\documentclass{article}

\everydisplay{\rm }

\begin{document}
\begin{center}

\LARGE The Volume Operator  \\ [.125in] for Spin Networks \\[.125in]
with Planar or Cylindrical Symmetry \\ [.25in] \large Donald E.
Neville \footnote{\large Electronic address: dneville@temple.edu}
\\Department of Physics
\\Temple University
\\Philadelphia 19122, Pa. \\ [.25in]
April 15, 2005 \\ [.25in]

\end{center}
\newcommand{\E}[2]{\mbox{$\tilde{{\rm E}} ^{#1}_{#2}$}}
\newcommand{\A}[2]{\mbox{${\rm A}^{#1}_{#2}$}}
\newcommand{\Np}{\mbox{${\rm N}'$}}
\newcommand{\Etwo}{\mbox{$^{(2)}\!\tilde{\rm E} $}\ }
\newcommand{\Etld }{\mbox{$\tilde{\rm E}  $}\ }
\def \ut#1{\rlap{\lower1ex\hbox{$\sim$}}#1{}}
\newcommand{\phst}{\mbox{$\phi\!*$}}
\newcommand{\psist}{\mbox{$\psi\!*$}}
\newcommand{\bea}{\begin{eqnarray}}
\newcommand{\eea}{\end{eqnarray}}
\newcommand{\be}{\begin{equation}}
\newcommand{\ee}{\end{equation}}
\newcommand{\nn}{\nonumber \\}
\newcommand{\rta}{\mbox{$\rightarrow$}}
\newcommand{\rla}{\mbox{$\leftrightarrow$}}
\newcommand{\eq}[1]{eq.~(\ref{#1})}
\newcommand{\Eq}[1]{Eq.~(\ref{#1})}
\newcommand{\eqs}[2]{eqs.~(\ref{#1}) and (\ref{#2})}
\large
\begin{center}
{\bf Abstract}
\end{center}
    This paper constructs a kinematic basis for spin networks with
planar or cylindrical symmetry, by exploiting the fact that the
basis elements are representations of an O(3) subgroup of O(4).
The action of the volume operator on this basis gives a difference
equation for the eigenvalues and eigenvectors of the volume
operator. For basis elements of low spin, the difference equation
can be solved readily on a computer, yielding the eigenvalues and
eigenvectors. For higher spins, I solve for the eigenvalues using
a WKBJ method. This paper considers only the case where the
gravitational wave can have both polarizations. The single
polarization case is considered in a separate paper. \\[.125in]
PACS categories:  04.60.Pp, 04.30.-w

\section{Introduction}

    This paper calculates the eigenvalues of the volume operator
for spaces with planar or cylindrical symmetry.  The volume
operator plays a key role in Thiemann's construction of a finite
Hamiltonian for spin network canonical quantum gravity
\cite{ThiemSN}

   The quantization of systems with planar or cylindrical symmetry
has been studied using both local field theory \cite{Kuc, Ashcyl,
MerMont, KorSam} and spin networks \cite{Nev, HusSmol}.  Before
one can construct a spin network Hamiltonian for these systems,
one must construct a volume operator. Bojowald has given a
discussion of some of the difficulties involved in constructing
this operator \cite{BojBH}.

    Systems with planar or cylindrical symmetry are the simplest where
gravitational wave propagation can occur.  Simpler systems, such
as homogeneous cosmologies and black holes with spherical
symmetry, have higher symmetry and fewer degrees of freedom, but
do not allow gravitational waves.  In a homogeneous cosmology
\cite{Bojcosm}, for example, every point is equivalent to every
other point. One may choose any one point, or vertex, of the spin
network as representative. The Hamiltonian changes intertwiners
and holonomies at that one vertex only. Since the action of the
Hamiltonian is purely local, there can be no propagation of
gravitational waves from point to point. For black holes with
spherical symmetry, with or without matter \cite{BojBH, TKBH,
AshBoj, Modesto, BojInh, BojSwi}, spherical symmetry rules out
gravity waves.

    The construction of a loop formalism typically proceeds in
two steps.  The states of the Hilbert space must satisfy seven
constraints.  In the first step one constructs a basis for the
Hilbert space which satisfies the six constraints which are
easiest to treat (Gauss and spatial diffeomorphism constraints).
The hardest constraint, the scalar or Hamiltonian constraint, is
left for the second step. The basis constructed in the first step
is the kinematical basis, and the dot product constructed in this
step is the kinematical dot product.  This structure is enough to
determine the eigenvalues of the volume operator.

    This paper completes the first step.  It constructs the
kinematical basis and dot product, and writes down equations which
allow the eigenvalues and eigenvectors of the volume operator to
be constructed numerically.  The paper does not go on to the
second step and construct a Hamiltonian.

    The rest of this paper proceeds as follows.  Section two describes
the topology of the spin network.  Section three defines the
volume operator and sets up the kinematical basis. Section four
writes out the eigenvalue equation for the volume operator in this
basis and derives an analytical solution for eigenvectors with
zero eigenvalue.  For non-zero eigenvalues, and for basis
functions of low spin, the equation can be programmed readily and
solved on a computer. Section five derives a WKBJ method to find
approximate eigenvalues of the volume operator.  This approach was
intended for high spins, but WKBJ surprises (as it often does). It
yields accurate eigenvalues even for basis functions of relatively
low spin.

    This paper does not impose the constraints which limit the
gravitational waves to a single polarization.  In classical
gravitational theory, and in quantum field theories of
gravitation, the single polarization case is easier.  However, in
the spin network case I found the single polarization case to be
harder.  I will leave a discussion of that case for a separate
paper.

\section{Topology of the Spin Network}

    A system with planar or cylindrical symmetry has two
commuting, spacelike Killing vectors. In the planar case, if one
shifts to the coordinates x,y suggested by the Killing vectors ,
the seven constraints simplify considerably, because derivatives
with respect to x and y may be dropped. (For the cylindrical case,
replace the planar coordinates x,y,z by coordinates $\phi$,z,r.
For simplicity in what follows, I shall discuss primarily planar
symmetry, and devote only an occasional remark to the cylindrical
case.) In particular, the x,y diffeomorphism constraints, and the
X,Y Gauss constraints simplify enough that they can be solved and
eliminated from the theory. Following Husain and Smolin, I fix
these four constraints by imposing the four gauge fixing
conditions \cite{HusSmol}\bea
    \E{x}{Z} &=& \E{y}{Z}= 0;\nonumber\\
    \E{z}{X} &=& \E{z}{Y}= 0.\label{Efix}
\eea Lower case letters x,y,z,\ldots denote global coordinates;
upper case X,Y,Z,\ldots denote local coordinates rotated by the
Gauss constraints. Setting the four constraints equal to zero and
solving yields four more equations, \bea
    \A{Z}{x} &=& \A{Z}{y}= 0;\nonumber\\
    \A{X}{z} &=& \A{Y}{z}= 0.\label{Afix}
\eea Both the triad and connection arrays are now block diagonal.
A 2x2 block contains fields with indices x,y and X,Y; a 1x1 block
contains the field with indices z,Z. (For cylindrical symmetry
substitute $\phi$,z,r and $\Phi$,Z, R for x,y,z and X,Y,Z.) The
SU(2) local gauge symmetry has been reduced to local U(1),
although the theory still contains all three generators $S_i$ of
su(2). In order to express the consequences of the U(1) symmetry
as clearly as possible, I choose the usual matrix representation
of su(2) where $S_z$ is diagonal. I suppose that the reduction in
components, \eqs{Efix}{Afix} has been carried out classically, and
I now set up a spin network formalism to quantize the reduced
theory.

     In classical general relativity, the theory allows one to choose
virtually any coordinate system; but it is not always clear which
coordinates lead to the simplest equations.  The choice of
coordinates is not obvious unless the system has a lot of
symmetry.

    Similarly in the spin network approach, the
choice for the topology for the network is not obvious unless the
system has a lot of symmetry.  I will arrive at the topology for
the planar case via the group theoretical approach favored by
practitioners in the field of quantum cosmology.  This approach
has been discussed extensively in the literature \cite{Kas,Bojth},
and for the most part my discussion will be  a summary of results.

    As a first step in constructing the symmetry reduced theory,
one uses group theory to construct a connection which embodies the
symmetry. Modern treatments manage to avoid anything so
\emph{d\'{e}class\'{e}} as solving the Killing equations, but in
the end results are the same. The group theory determines the
number of independent components and their form. It also
determines the support of the connection fields. (In effect this
step has been carried out already at \eqs{Efix}{Afix}, where we
determined the number of independent components, and determined
their support to be the z axis.)

       However, group theory by itself does not supply the
topology of the spin network. For example, in a homogeneous
cosmology, every point of the space is equivalent to every other
point.  Not suprisingly then, the group theory predicts  support
for the homogeneous connection is limited to a single point. We
know that, in the full theory, each vertex is connected to some
number of edges. In effect, the group theory supplies a vertex
(the point) but no edges. Motivated by the full theory, one
introduces edges, and promotes the connections to holonomies
integrated over these edges.

    In the cosmological case, one can justify the introduction of
edges \emph{\'{a} posteriori} by constructing a Bohr compactified
Hilbert space which has some very unusual and useful properties
\cite{ABL}. But it is unlikely anyone would have thought of doing
this, had they not been motivated by the existence of edges and
holonomies in the full, non-symmetric, theory.

    Similarly, in the planar case, the group theory predicts that
support for the connections is the z axis, but produces no edges.
Motivated by the full theory, one introduces vertices along this
axis, as well as two edges radiating from each vertex in x and y
directions. The connections are then promoted to holonomies
integrated over the edges.

    Holonomies on the z axis look like holonomies in the full
theory.  Each holonomy is integrated from one vertex to the next.
Holonomies on the x and y edges are treated differently.  These
edges are given the topology of a circle; the two ends of each
edge are identified.

    This identification is as first sight somewhat puzzling. (Most
of the rest of this section records my initial doubts, and
eventual acceptance of circular topology for the xy edges. Readers
who are comfortable with this choice of topology may wish to skip
to the last three paragraphs of this section.) One would like to
think of the symmetric theory as a reduction of a full theory.
That is, whenever one has a symmetry, one expects to start from
the action of the full Hamiltonian on the full space, and end with
the action of a simpler Hamiltonian on a smaller space. The
ultimate example of such a reduction is a homogeneous cosmology,
where one starts from a Hamiltonian acting on every point, and
ends with a simpler Hamiltonian acting on a space consisting of a
single, representative point.

    It is therefore natural to visualize the full spin network as
constructed from a basic unit, or module, which is repeated over
and over to create the full space. For the homogeneous cosmology
the basic module would be a unit cell containing only a single
vertex, plus three edges. For the edges one could choose (say)
three edges extending in positive coordinate directions, with the
midpoint of each edge at the vertex. The basic cell would then be
six half-edges radiating from a central point.  Repeating this
cell over and over generates the full space. (If the system
possesses additional symmetries such as isometry, one may need
fewer than three edges.  A change in the number of edges does not
affect the present argument, and I will ignore this possibility.)

    To return to my point: the modular picture does not immediately
suggest an $S_1$ topology for each of these edges.  Each basic
cell contains an intertwiner at the vertex, plus six holonomies
associated with the half-edges. When the full space is obtained by
multiplying these basic units together, one index on each holonomy
is already contracted with an index on an intertwiner; as for the
remaining index, one might expect to contract it with the
corresponding index on a neighboring cell. It is not obvious one
should contract this index with another index in the same cell,
which is what happens when one identifies ends and imposes an
$S_1$ topology on each edge.

     If one does not identify ends and contract indices, each
cell would have six "dangling" SU(2) indices, indices not
contracted with any other index. Those indices can be ignored when
Gauss-rotating the cell, however.  The uncontracted indices may be
viewed as merely an artifact of splitting the full space up into
identical cells.

    However, the "dangling index" picture does not hold up well
when we consider reduction from the full to the symmetry reduced
theory; circular topology seems essential. In more detail (and
continuing with the homogeneous example for simplicity), label
each cell by its vertex v and label each half edge e by an index
$j_e$, which denotes the total spin of the holonomy on that edge
(if one is using 2$j_e$ + 1 dimensional irreducible
representations of SU(2) on each edge); or $j_e$  denotes the
number of spin 1/2 holonomies (if one is using a product of spin
1/2 matrices on each edge).  By homogeneity, the two halves of a
given edge must have the same $j_e$.  Then the spin network
wavefunctional for the full space is \be
    \psi = \Sigma_{je} \Pi_v \psi_v ({j_e})_m.\label{fullpsi}
\ee $\psi_v$ denotes the wavefunctional of a single cell.  The
full space is a product of such cells (the $\Pi$); and one must
sum over possible assignments of edge spins (the $\Sigma$).  The
subscript m is shorthand for the six SU(2) indices on $\psi_v$.
These indices are contracted with corresponding indices on
neighboring cells.

    Since every point is equivalent to every other, the modules must
identical, initially, and the dynamics must keep them identical.
Dynamics consists of identical transformations applied to every
cell. In this spirit, I write the Hamiltonian constraint as a sum
of constraints, one for each unit cell.\[
    H = \Sigma_v H_v.\]
The constraint 0 = H$\psi$  implies \be
    0 = \Sigma_{je} H_v \psi_v({j_e})_m[ \Pi_{v'\neq v}\psi_{v'}({j_e})]_m
    \label{fullH}\ee
I would like this to reduce to  \[
    0 = \Sigma_{je}H_v \psi_v({j_e})_m,\]
i.e. to a single term in H acting on a single cell of $\psi$.
However, this will not happen, in general, because the m indices
on the square brackets in \eq{fullH} have ranges which depend on
the $j_e$. The square brackets therefore do not factor out of the
sum over $j_e$.

    I can, of course, eliminate the sum over $j_e$ by confining
myself to a single set of $j_e$; but this is too restrictive. I am
left with the standard choice in the literature: identify opposite
ends of each edge; trace over indices at opposite ends.  Each cell
then has no SU(2) indices to contract with neighboring cells.  The
square brackets in \eq{fullH} are now identical, and they factor
out of the sum. Equivalently, the traces change $\psi$,
\eq{fullpsi} to\be
     \psi \rta \Pi_v \Sigma_{je} \psi_v ({j_e}),\label{newfullpsi}
\ee where now there are no SU(2) indices on $\psi_v$.  Each cell
$\psi_v ({j_e})$ is an SU(2) scalar.

    I treat the planar case in the same way as the homogeneous
case just discussed.  The basic module is now the z line, plus two
x and y edges at each vertex, with midpoints located at the
vertex. The transverse edges are given the $S_1$ topology. For
example, if I write out only the x edges and x indices at a given
vertex,\[
    \psi_v = \cdots h[A_x, e_{xi}/2]_{m',
    mi}I_{mi,mf}h[A_x,e_{xf}/2]_{mf, m'}\cdots \]
Here $e_{xi}$/2, for example, is the half-edge entering the vertex
with intertwiner I.  This structure may be rewritten in a manner
which hides the traced indices.\[
    \psi_v = \cdots I_{mi,mf}h[A_x,e_{xf}]_{mf, mi}\cdots .\]
Those indices usually play no dynamical role anyway, since the
volume operator acts at vertices, i.e. its action affects only the
$m_i,m_f$ indices.

    Loosely, one can describe the rewritten holonomy as a "loop"
holonomy which leaves from and returns to the same vertex. I will
sometimes use this way of speaking, but note the holonomy is
integrated over a line, not over a loop in the xy plane.  The
holonomy depends only on the connection $A_x$.

    The final modular structure allows communication in the
longitudinal direction, but no communication in transverse
directions. Gravitational waves can propagate only along z.

\section{Spin Network Basis States}

    In the previous section I proposed a spin network topology
consisting of an infinite line dotted with vertices; at each
vertex two edges extend in the positive x and y directions with
ends identified.  I now associate a holonomy with each segment of
the line and each transverse edge. For example, for the transverse
x edge, \be
    h[A_x, e_{xf}] = \exp \int i \A{B}{x}S_{B} dx. \label{hol}
\ee The index B ranges over X and Y only.  The expression on the
right is not a trace.  The h$[A_x, e_{xf}]$ matrix has two SU(2)
indices.  Both are contracted with  corresponding indices on the
intertwiner at the vertex; I am using the "loop" viewpoint
described in the last three paragraphs of the previous section.
The holonomies along z edges are similar to those in \eq{hol},
except that the su(2) valued connection is \A{Z}{z}$S_Z$ and the
integration is from one vertex to the next.

    The next step is to determine the selection rule which the
holonomies must obey, because they preserve the residual gauge
symmetry U(1) at each vertex. U(1) transformations are generated
by the surviving Gauss constraint,
 \be
    G[\Lambda] = \int \Lambda (\partial_z \E{z}{Z} + \epsilon_{ZAB}
            \A{A}{a}\E{a}{B}).\label{Gauss}
 \ee
 A typical vertex, located at coordinate z, will have
 two z-edge holonomies, one beginning
 and one ending at z. The constraint \eq{Gauss} generates the
 infinitesimal transformation  $\A{Z}{z}\rightarrow
 \A{Z}{z} - \partial_z \Lambda $ which leads to the following
 finite transformation of the z holonomies.
 \bea
    H[A_z]&:=& h[A_z, e_{zi}]h[A_z, e_{zf}]\nn
            &=&\exp(i\int^z m_{zi} \A{Z}{z} dz) \exp (i\int_z m_{zf} \A{Z}{z} dz) \nonumber \\
            &\rightarrow& \exp[- i\Lambda (z) (m_{zf} -
        m_{zi})]H[A_z].\label{hzdef}
 \eea
 Here I have replaced $S_Z \rightarrow m_{zf}$ or $m_{zi}$; there is no point
 to retaining the $S_Z$, since U(1) does not mix
 different eigenvalues of $S_Z$.  $m_{zf}$ and $m_{zi}$ are integers or half-integers.
 f and i label the outgoing and incoming z edges, respectively beginning and
 ending at the vertex.

    The same vertex will have two transverse edge holonomies beginning at the
 vertex.  For the transverse connection $A_x$, the infinitesimal
 transformation is
 \[
        \A{C}{x} S_C \rightarrow \A{C}{x} S_C + [\A{C}{x}
 S_C,i \Lambda S_Z],
 \]
 which implies the following finite transformation, when the
 connection is promoted to a holonomy
 \bea
    h[A_x, e_{xf}]_{mf'mf}&\rightarrow&
        (\exp[-i\Lambda S_Z]h[A_x, e_{xf}]\exp[+i\Lambda S_Z])_{mf'mf}\nonumber\\
    &= & \exp[-i\Lambda (m_{f'}-m_f)]h[A_x,
    e_{xf}]_{mf'mf},\label{hxdef}
    \eea
 and a similar equation for $h[A_y, e_{yf}]_{nf'nf}$.

    Collecting together the x, y, and z contributions, I find that a U(1)
 transformation multiplies a vertex by the overall phase
 \be
    \exp[-i(m_{zf} - m_{zi} + m_{f'} - m_f + n_{f'} - n_f)\Lambda(z)].\label{U1phase}
 \ee
 U(1) invariance requires the relation
 \bea
    2F&:=& m_f-m_{f'} + n_f-n_{f'}\nonumber\\
        &=& m_{zf} - m_{zi} \label{Frule}
 \eea

    Note that the intertwiner at each vertex no longer has to be a
product of SU(2) 3J symbols.  The intertwiner can be any product
of Kronecker deltas assigning specific values to the m's and n's,
provided  the selection rule \eq{Frule} is obeyed at the vertex.

    It is straightforward  to concoct a kinematic Hilbert space
for the z holonomies, \eq{hzdef}.  The functions exp[i m $\theta$]
constitute a complete set of functions periodic on $[0,4\pi]$, for
m integer or half-integer.

     However, I cannot use the SU(2) Haar measure to
formulate a dot product for the transverse edge holonomies.  To
see this, I construct the simplest edge holonomy, that given by
the spin 1/2 representation of h. Denote the axis of rotation by
$\hat{n}$. Since $\hat{n}$ must lie in the XY plane, it has the
form   \be
        \hat{n} = (\cos \beta, \sin \beta,0).\label{n}
 \ee
for some angle $\beta$.  Then the spin 1/2  holonomy becomes \bea
    h^{(1/2)} &=& \exp[i \hat{n} \cdot \sigma
\theta/2]\nonumber\\
    &=& \cos(\theta/2) + i\hat{n}\cdot \sigma
    \sin(\theta/2)\nonumber\\
    &=& \exp[-i\sigma_z(\beta-\pi/2)/2]\exp[i\sigma_y
    \theta/2] \nonumber \\
           &\times& \exp[+i\sigma_z(\beta-\pi/2)/2].\label{dhalf}
\eea On the last two lines I have written the usual Euler angle
decomposition for this rotation.  It is clear that the rotation
with axis confined to the XY plane depends on only two Euler
angles, rather than the generic three.  Let $h^j $ denote the 2j+1
dimensional matrix representation of SU(2).  Then integration over
only two angles
\[
    \int (h^{j'*})_{a'b'} (h^{j})_{ab} \sin \theta
    ~d\theta ~d\beta
\]
 guarantees only $a-b = a'-b'$ but not the stronger constraints
$a' = a$, $b'= b$; the latter constraints must be satisfied in
order for the $\theta$ integration to yield $j' = j$.

    So far, I have been trying to follow a logical order: first
construct an orthonormal basis; later, investigate the action of
the volume operator on this basis.  Logical order has
 not revealed a suitable dot product for the loop holonomies.

    It turns out it is better to reverse logical
 order.  Start by investigating the action of the volume operator
(on the  spin 1/2 holonomy constructed above). This action
suggests a basic structure which can be used to build higher order
polynomials in the matrix elements of the spin 1/2 holonomy, while
maintaining simple behavior under the action of the volume
operator.  Finally, construct the  kinematic dot product suggested
by the structure of these polynomials.

    I will need to define the action of the volume operator (more precisely
the square of the volume operator). This operator is the product
 \be
    (V_3)^2 = \epsilon_{ZBC}\E{z}{Z}\E{x}{B}\E{y}{C}.\label{V}
    \ee
Each $\tilde{E}$ operator must be integrated over an area, in
order to make the volume invariant under spatial diffeomorphisms.
In particular \E{x}{B} must be integrated over an area in the yz
plane. For clarity I have suppressed these areas in \eq{V}.
However, I cannot ignore them completely.  Their precise extent is
needed in order to determine the location on the yz area where
\E{x}{B} can find an \A{B}{x} to grasp.

    That location is easy to find in the context of the full
theory. The \A{B}{x} field has its support on an x edge, whereas
the \E{x}{B} field is confined to the yz plane. The yz plane
intersects the x edge only at the vertex. Therefore \E{x}{B} acts
where the x edge holonomy meets a vertex.  In the reduced theory,
this means \E{x}{B} acts at the two ends of the holonomy,
\eq{hol}, where the holonomy intersects the vertex.

    I choose the yz area narrow enough in the y direction that
the area includes only one vertex.  When the full theory reduces
to the symmetric one, neighbors in the y direction disappear
anyway.

    In the z direction, I could choose an area which overlaps two or
more vertices.  Then \E{x}{B} could act on one  vertex at one z,
while the other two triads in the volume operator act on a
different vertex at a different z. I assume the yz area may extend
in the z direction halfway to the next vertices, but not all the
way, so that the volume operator can grasp lines exiting from only
a single vertex. Equivalently, I assume all three triads grasp
edges exiting from the same vertex. This assumption has the
advantage of simplicity. It is also reasonable, since propagation
(which demands operators that change more than one vertex) is
associated with components of the Riemann tensor, not with the
volume.

    The $\tilde{E}$ operators in a canonical quantization act like
functional derivatives with respect to the corresponding A
operators.  When acting on an edge holonomy, the $\tilde{E}$ bring
down an su(2) generator at each end of the holonomy (never in the
middle, because the volume operator acts only where the holonomy
meets a vertex). For example the x holonomy is replaced by the
following anticommutator. \bea
    \E{x}{A} h[A_x, e_{xf}] &=& \E{x}{A} \exp[i\int \A{B}{x} S_B
    dx]\nonumber \\
                      &=& (1/2)[S_A h[A_x, e_{xf}]+
                      h[A_x, e_{xf}] S_A)]\gamma\kappa.\label{grasp}
                      \eea
Subscripts A,B = X, Y only.  I have omitted the triple delta
functions, because they are always cancelled by the area and line
integrals associated with \E{x}{A} and \A{B}{x}.  The overall
factor $1/2$ is a relic of the integrals over the delta's, $1/2$
because the deltas occur at the endpoints of the edge integration.
$\gamma\kappa$ is the product of Immirzi parameter times 8$\pi$ G
; $\hbar = c = 1$.

    The anticommutator means that the volume operator generates an
infinitesimal O(4) transformation.  To see this, specialize
\eq{grasp} to the spin 1/2 case: $h[A_x, e_{xf}]\rta h[A_x,
e_{xf}]^{(1/2)} $ (see \eq{dhalf}); and $S_A \rta \sigma_A /2$.
Now consider the following two sets of SU(2) transformations,\bea
    R := \{U^{-1} h[A_x, e_{xf}]^{(1/2)}U\};\nn
    B :=\{ U h[A_x, e_{xf}]^{(1/2)}U\}.
    \label{boost}\eea
The R's are just ordinary rotations; their infinitesimal form is a
commutator.  The infinitesimal form of the B's, however, is an
anticommutator. The sets R and B are special cases of the set of
transformations \{$U'h[A_x, e_{xf}]^{(1/2)}U$, $U'\neq U$\}.  The
set is SU(2)$\bigotimes$SU(2).

    Modulo fine points about covering groups, SU(2)$\bigotimes$SU(2)
is O(3)$\bigotimes$O(3)= O(4).  To exhibit the O(4) structure,
introduce the suggestive notation\be
    i h[A_x, e_{xf}]^{(1/2)} := \left\lbrack\matrix{
                     x_3 + i x_4 & x_1 - i x_2 \cr
                     x_1 + i x_2 & -x_3 + i x_4 \cr}
                        \right\rbrack
     \label{O4matrix}\ee
Since the matrices $U'$ and $U$ are unimodular, the transformation
$U'h[A_x, e_{xf}]^{(1/2)}U$ preserves the determinant, which is
\[
    -(x_1^2 + x_2^2 + x_3^2 + x_4^2).\]

    The  transformations R turn out to be ordinary rotations;
they leave $x_4$ invariant.   A transformation B which has axis
along direction i rotates $x_i$ into $x_4$ ($1\leq i \leq 3$).
(The B stands for boost. The B's are of course rotations, not
Lorentz boosts, but they become Lorentz boosts when their angular
parameter is continued to a pure imaginary value.)

    If possible, I would like to avoid using O(4) spherical
harmonics as a kinematic basis.  O(4) is double trouble: the
harmonics are products of two rotation matrices, and the
Clebsch-Gordan coefficients are products of two SU(2)
Clebsch-Gordan coefficients.  Fortunately, it is possible to use
harmonics of an O(3) subgroup of O(4).

    To identify the O(3) subgroup, I calculate the components $x_i$ of
$h[A_x, e_{xf}]^{(1/2)}$, \eq{O4matrix}.  The following traces
give components $x_1$ through $x_4$. \bea
    x_1+ i x_2 &=& Tr[ih[A_x, e_{xf}]^{(1/2)}\sigma_+]/\sqrt{2}\nn
                &=& -\sin(\theta/2)\exp[i\beta];\label{pl}\\
    i x_4 &=& Tr[ih[A_x, e_{xf}]^{(1/2)} 1]/2 \nn
            &=& i\cos(\theta/2);\label{one}\\
    x_1 - i x_2 &=& Tr[h[A_x, e_{xf}]^{(1/2)}\sigma_-]/\sqrt{2}\nn
                &=& -\sin(\theta/2)exp[-i\beta];\label{min}\\
    x_3 &=& Tr[ih[A_x, e_{xf}]^{(1/2)}\sigma_z]/2 \nn
        &=& 0.\label{z}
\eea I have introduced the usual raising and lowering operators
\be
    \sigma_{\pm} = (\sigma_x \pm i \sigma_y)/\sqrt{2}.\label{sigpm}
\ee  In computing the above traces, it is convenient to write
\eq{O4matrix} as $ \vec{x}\cdot\vec{\sigma} + i x_4$.

    From \eq{z}, the component $x_3$ vanishes. This is not an
accident.  There is no $\sigma_3$ in the expansion of \eq{dhalf}
because the rotation is confined to the XY plane. Further, $A_z$
is the only $S_z$ valued connection left in the theory, after the
gauge fixing, \eq{Afix}; and neither the Gauss constraint nor the
volume operator have the power to change $A_z$ into an $A_{x,y}$.
Hence the relevant operators maintain $x_3 = 0$. This suggests we
do not need the full O(4), but only the little group which leaves
invariant the vector \be
    (x_1,x_2,x_3,x_4) = (0,0,1,0). \nonumber \ee

    The following theorems will make this idea more precise.  First
I recall some standard O(4) theory in order  to identify the
generators of boosts and rotations. For the SU(2)$\bigotimes $
SU(2) transformation $U'h[A_x e_{xf}]U$, let $s'$ and $s$ denote
the generators of $U'$ and $U$ respectively, so that
\[
    [s'_i,s'_{j}] = i \epsilon_{ijk}s'_k ; \]
\[    [s_i,s_{j}] = i \epsilon_{ijk}s_k ; \]
\be     [s'_i,s_{j}] = 0 .\label{O4comm} \ee

Then the generators of boosts and rotations are given by \bea
    b_i &=& (s_i + s'_i)/2;\nn
    r_i &=& (s_i - s'_i)/2. \label{brdef}\eea
Proof: write $s$ and $s'$ as \bea
    s &=& (s+s')/2 + (s-s')/2;\nn
    s'&=& (s+s')/2 - (s-s')/2.\eea
Demand that the general transformation $U'h[A_x, e_{xf}]U$ reduce
to the special transformations defined in \eq{boost}.  For the
boosts, the matrices $U'$ and $U$ must have the same generator,
which means only $(s+s')$/2 can contribute; and similarly for the
rotations, only $(s-s')$/2 can contribute.

    Next I need to identify the little group.  Let $(r_1,r_2,r_3)$
be generators of su(2) and let $\vec{V}$ transform like a vector
under the $r_i$, that is \[
    [r_i,r_j] = i\epsilon_{ijk} r_k;
            [r_i,V_j] = i\epsilon_{ijk}V_k.\]
Then\bea
    (r_1,r_2,r_3) &\cong & (b_1,b_2,r_3);\nn
    (V_1,V_2,V_3) &\cong & ( -x_2,+x_1,x_4).\label{isomorph} \eea
That is, the generators  $(b_1,b_2,r_3)$ of o(4)have the same Lie
algebra as o(3); and the quantities $( -x_2,+x_1,x_4)$ rotate like
a vector under the action of these generators.  Note that $x_3$
does not rotate; $(b_1,b_2,r_3)$ generate the little group.

    On the second line of \eq{isomorph}, the subscripts 1 and 2 have
been interchanged, equivalent to a ninety degree rotation of the
V's. The rotation is necessary because $b_1$ (for example) is not
quite $r_1$. $b_1$ rotates $x_1$ into $x_4$, whereas $r_1$ rotates
$V_2$ into $V_3$. A rotation is needed to exchange $V_2$ and $V_1$
before the isomorphism will work.  For later convenience I rewrite
this isomorphism in terms of eigenstates of the ninety degree
rotation:\be
    (V_{\pm},V_3) \cong  (\pm i x_{\pm},x_4).\label{pmisomorph} \ee

    Proof of \eq{isomorph}:  it is straightforward to verify that
the $(b_1,b_2,r_3)$ obey the Lie algebra of o(3), by using the
definitions \eq{brdef} of the b's and r's, together with the
commutation relations \eq{O4comm} for the $s$ and $s'$. To verify
that the x's rotate like a vector, note that the $U$ in $U'h[A_x,
e_{xf}]U$ is $\exp[(i(b_1,b_2,r_3)\cdot \vec{\alpha}]$, while the
$U'$ is the same, except for $\alpha_3\rightarrow - \alpha_3$.
Therefore for the $b_i$, the infinitesimal transformation is an
anticommutator (with $\sigma_i /2$, since $h[A_x, e_{xf}]$ is spin
1/2); for $r_3$ the infinitesimal transformation is a commutator.
$b_1$ (for example) generates the following infinitesimal
transformation $\delta [i h[A_x, e_{xf}]^{(1/2)}$:\be
    [i h[A_x, e_{xf}]^{(1/2)},i\sigma_1/2]_+ = \left\lbrack\matrix{
                     i x_1 &  -  x_4 \cr
                     -x_4 & i x_1 \cr},
                        \right\rbrack
\ee or\be
        (\delta x_1,\delta x_2,\delta x_3,\delta
        x_4)=(-x_4,0,0,+x_1).\label{b1transf}\ee
Compare this to $[i r_i,V_j]_- = -\epsilon_{ijk}V_k$, or\[
    \delta V_2= - V_3;  \delta V_3 = + V_2.\]
After relabeling these equations as required by \eq{isomorph},
they become \eq{b1transf}. One proceeds in similar manner to prove
the rest of the isomorphism.

    I will refer to the vector on the right in \eq{pmisomorph}
or \eq{isomorph} as the basic vector.  By inserting \eq{pl}
through \eq{min} into \eq{pmisomorph}, one can express the basic
vector in terms of the angles $\beta$ and $\theta$. \bea
    (V_{\pm},V_3) &\cong & (\pm i x_{\pm},x_4)\nn
                &=& (\sin(\theta/2)\exp[\pm i (\beta - \pi/2)],
                        \cos(\theta/2))\label{unitv}\eea
Evidently the basic vector is a unit vector.  Also, we can
construct Condon-Shortley spherical harmonics $Y_{1m}$ from the
components of the basic vector \cite{CondShort}, but these
spherical harmonics will have non-standard arguments:
$Y_{1m}(\theta/2,\beta -\pi/2)$ rather than the usual
$Y_{1m}(\theta,\beta )$.

    The next step is to build up more complex holonomies at the vertex:
multiply together L matrix elements of $h[A_x, e_{xf}]^{1/2}$ to
form all possible homogeneous polynomials of order L in the matrix
elements.  These polynomials form a rank L reducible
representation of the little group O(3).  Since we are multiplying
together L identical vectors, we must break into irreducible
representations by symmetrizing and taking traces;
antisymmetrization gives zero.  The irreducible representations
are then just the $Y_{L'm}$ with $L' = L, L-2, L-4$,\ldots.

    Therefore the set \{$Y_{Lm}(\theta/2,\beta -\pi/2)$\} is complete,
and I adopt it as a basis. Because these $Y_{Lm}$ have argument
$\theta/2$ rather than $\theta$, the usual dot product for the
spherical harmonics \[
    \int_0^{2\pi}d\beta \int_0^{\pi} \sin \theta ~d\theta\]
will have to be modified slightly. \be
    <L'm'\mid L m> = \int_0^{2\pi} d \beta
        \int_0^{2\pi}\sin(\theta/2)~d(\theta /2)~
        Y_{L'm'}^* Y_{L m}. \label{dotpr}
\ee

    I should perhaps emphasize that I have actually
constructed two bases and two dot products, one for x holonomies
with angles $\theta_x, \beta_x$, and another for y holonomies with
angles $\theta_y, \beta_y$. These angles are unrelated except in
the one-polarization case, where classically, the two axes of
rotation are at right angles, $\beta_x = \beta_y \pm \pi/2$. In
that case, the kinematic Hilbert space and dot product must be
reconsidered from scratch.

    Note that the $Y_{Lm}$ basis is far easier to use than (say) the set
$\{\mathcal{D}^j(-\beta + \pi/2, \theta,\beta - \pi/2)_{0m}\}$,
which would be a straightforward generalization of \eq{dhalf} to a
rotation matrix of higher spin. The $\mathcal{D}^j_{0m}$ are
orthonormal, but they are not representations of the O(3) little
group, and the set $\{\mathcal{D}^j_{0m}\}$ is not closed under a
grasp by the volume operator.

    As for the $Y_{Lm}$, a grasp by \E{x}{+}, for example, multiplies
 the basic vector by the spin 1 representation of $S_+$; this basic map
induces a map of the higher harmonics into themselves, a map given
by the spin L representations of $S_+$. Symbolically, \be
    \E{x}{+}Y_{Lm} = \Sigma_{m'} Y_{Lm'}<L,m'\mid S_+\mid L,m>,
 \label{+transf}\ee
 with \cite{CondShort}\be
    <L,m\pm 1\mid S_A = S_{\pm}\mid L,m> = \sqrt{(L \mp m)(L \pm m +
    1)/2}.\label{Spm}
\ee I have now completed the construction of the kinematic basis
and dot product; and described the action of the volume operator
triads on this basis.

\section{Eigenvalue Equation for the Volume Operator}

    To summarize the results of the previous section: I now have a
basis set of holonomies; at each vertex there are two z-edge
holonomies \be
   H[\A{Z}{z}]= \exp[i\int^z m_{zi} \A{Z}{z} dz] \exp [i\int_z m_{zf} \A{Z}{z} dz]
    \label{zhol}\ee

plus two "loop" holonomies \be
    Y_{L_x m_x}(\theta_x/2,\beta_x -\pi/2) Y_{L_y m_y}(\theta_y /2,\beta_y
    -\pi/2).\label{loophol}\ee

    At \eq{Frule} I worked out the conseqences of the surviving U(1)
gauge invariance, but for standard holonomies $h_{m'm}$ rather
than the new $Y_{Lm}$ basis.  It is easy to see from \eq{n} that a
gauge rotation changes only $\beta$ while leaving $\theta$ alone;
therefore to discover the gauge behavior of the Y's we need to
study their $\beta$ dependence. The holonomies $h_{m'm}$ have
$\beta$ dependence $\exp[i(m-m')\beta]$ (see, for example,
\eq{dhalf}). Except for a normalization, $Y_{L_x m_x}$ is the
rotation matrix $\mathcal{D}^{Lx}_{0mx}$, which has $\beta$
dependence $\exp[i m_x \beta]$ Therefore at \eq{Frule} the
differences m - m' should be replaced by $m_x$; and there is a
similar replacement for the y indices. The U(1) selection rule,
\eq{Frule} simplifies to \bea
    2 F&=& m_x + m_y\nonumber\\
        &=& m_{zf} - m_{zi} .\label{Fnew}
 \eea

    The z holonomies of \eq{hzdef} are eigenfunctions of the $\E{z}{Z}$
 factor in the volume operator; and the remaining $\tilde[E]$ operators
 in the volume map the loop holonomies \eq{loophol} into
 themselves as at \eq{Spm}  Therefore the volume operator
  will not change $m_{},m_{zf},L_x$,or $L_y.$  It will not change F$= (m_x +
  m_y)/2$, F for "fixed", because of the selection rule \eq{Fnew}.
 The volume operator can change the quantity \[\mbox{D} := (m_x
- m_y)/2, \] D for "difference". Therefore an eigenfunction of the
volume operator will be the product of the two z-edge holonomies
\eq{hzdef} times a sum \be
    \mid \lambda;L_x L_y F> = \Sigma_D Y_{L_x m_x}Y_{L_y m_y}
    c(m_x,m_y)\label{veigen}\ee
For simplicity I have suppressed the L and F dependence of the
c's.  The z dependence of the volume operator \eq{V} acts on each
of these basis elements as follows. \bea
    (V_3)^2 H[A_z]\mid \lambda;L_x L_y F>& &\nonumber\\
    &=&(\gamma\kappa/2)(m_{zi}-m_{zf})H[A_z] \nonumber \\
     & \times&   (V_2)^2\mid \lambda;L_xL_y F>.\label{V3}
\eea $H[A_z]$is defined at \eq{hzdef}; the constants
$\gamma\kappa/2$ are as at \eq{grasp}. $(V_2)^2 $ is the
determinant of the 2x2 subblock.

    To complete the action of $V_3$, \eq{V3}, I must determine how
$V_2$ acts on the expansion \eq{veigen}. I shift from X,Y to the
 combinations X $\pm$ iY in $V_2$, to
simplify later matrix elements. \bea
    \E{a}{\pm} &:=& (\E{a}{X} \pm i \E{a}{Y})/ \sqrt{2};\nonumber\\
    (V_2)^2 &=& \epsilon_{ZAB}\E{x}{A}\E{y}{B} \nn
            &=&\epsilon_{Z-+}\E{x}{+}\E{y}{-}+
                    \epsilon_{Z+-}\E{x}{-}\E{y}{+}\nonumber\\
        &=& i (\E{x}{+}\E{y}{-}- \E{x}{-}\E{y}{+}).\label{V2def}
\eea The i and the minus sign come from $\epsilon_{Z \mp \pm} =
\pm i$. Note how plus indices are always contracted with minus
indices. For example,\be
    \E{x}{X} \sigma_X + \E{x}{Y}\sigma_Y = \E{x}{+} \sigma_- + \E{x}{-}\sigma_+
    \ee
\E{x}{+} is essentially the functional derivative with respect to
 \A{-}{x}, therefore when it acts on
 the basic holonomies (\eq{dhalf}, and \eq{pl} thru
 \eq{min}), the \A{-}{x} functional derivative replaces each
 holonomy by its anticommutator with $\sigma_+$. (Again, plus
indices always pair with minus.) The basic holonomies transform as
a spin 1 representation under this anticommutation.  When acted on
by \E{x}{\pm}, therefore, the Y's transform as the 2L+1
dimensional representation of $S_{\pm}$. \bea
    (V_2)^2\mid \lambda;L_x L_y F> &=& i(\gamma\kappa/2)^2
    \nonumber \\
         & \times& \Sigma c(m_x,m_y)[Y_{L_x m_x+1}<L_x m_x+1\mid S_+ \mid L_x
                    m_x> \nonumber\\
            &\times &Y_{L_y m_y-1}<L_y m_y-1\mid S_- \mid L_y m_y> \nonumber\\
            & -& (x\leftrightarrow y)]\nonumber \\
            &=& (\gamma\kappa/2)^2 \lambda \mid \lambda;L_x L_y
            F>\label{V2}
\eea On the last line I have assumed the state is an eigenstate of
$(V_2)^2$.  Multiplying both sides by $Y_{L_x  m_x} Y_{L_y m_y}$,
using the kinematic dot product introduced at \eq{dotpr}, and the
matrix elements given at \eq{Spm}, I find the following eigenvalue
equation for $\lambda$. \bea
    2\lambda ~c(m_x,m_y)&=&
    i g(L_x,m_x,L_y,m_y)~c(m_x-1,m_y+1)\nonumber \\
        &-& i g(L_y,m_y,L_x,m_x)~c(m_x+1,m_y-1);
                        \nonumber \\
    g(L_x,m_x,L_y,m_y)&=&\sqrt{(L_x-m_x+1)(L_x+m_x)}\nonumber\\
        &\times& \sqrt{(L_y+m_y+1)(L_y-m_y)}.\label{recurr} \eea

In terms of $\lambda$ the eigenvalues $ \lambda_3$ of the original
operator $(V_3)^2$ are, from \eqs{V3}{V2}, \be
    \lambda_3 = (\gamma\kappa/2)^3 (m_{zi}-m_{zf}) \lambda
    \label{lmbda3}
\ee

    Because F$= (m_x+ m_y)/2$ is held fixed, only the quantity
\be
    D := (m_x-m_y)/2 \label{Ddef}
\ee is incremented in \eq{recurr}.  The equation is an ordinary
difference equation masquerading as a partial difference equation.
To exhibit the ordinary difference character, I make the following
replacements \bea
    m_x &=& F + D;\nonumber \\
    m_y &=& F-D; \nonumber \\
    c(m_x\pm 1,m_y \mp 1)&=& c(D \pm 1).
\eea With these replacements, \eq{recurr} becomes \bea
    2 \lambda ~c(D)&=&i\sqrt{(L_x-F-D+1)(L_x+ F + D)}\nonumber \\
        &\times&
        \sqrt{(L_y+F-D+1)(L_y-F+D)}~c(D-1)
 \nonumber \\
    &-&i \sqrt{(L_x+F+D+1)(L_x- F - D)} \nonumber \\
        &\times&\sqrt{(L_y+F-D)(L_y-F+D+1)}~c(D+1)\label{recurrD}
 \eea
    I have been unable to find a compact, analytic solution to
the above equation. The next three paragraphs, which describe my
efforts to find such a solution, are perhaps of interest only to
readers who are already familiar with the standard literature on
ordinary difference equations \cite{MT}. Some readers may wish to
skip these paragraphs on first reading.

    To make contact with the standard literature, which treats
primarily equations with rational coefficients, I need to get rid
of the square roots .  This is easily done by a change of
dependent variable. \bea
    c(D) &=& \sqrt{(L_x+m_x)! /(L_x-m_x)!}\nonumber\\
          &\times&      \sqrt{(L_y+m_y)!/(L_y-m_y)!} ~d(D)
                     \nonumber \\
        &=& \sqrt{(L_x+F+D)!/(L_x-F-D)!}\nonumber \\
        &\times& \sqrt{ (L_y+F-D)!/(L_y-F+D)!} ~d(D)\label{ddef}
\eea If this is inserted into \eq{recurr} and square roots
cancelled, the resulting equation for d is \bea
    2\lambda ~d(D)&=&i (L_y+F-D+1)(L_y-F+D)~d(D-1) \nonumber \\
            &-&i (L_x+F+D+1)(L_x-F-D)~d(D+1).\label{deq}
\eea The coefficients are now rational, but they are quadratic
functions of the independent variable D.  Solutions when the
coefficients are linear are already available in the literature,
but for quadratic coefficients one must construct a solution.

    One approach is to assume the solution is a series of factorials
d(D) =$ \Sigma_n a_n /(D-n)!$. (Factorials play the same role in
the theory of difference equations that powers do in the theory of
differential equations.)  The problem of determining the d's is
turned into the problem of determining the coefficients $a_n$; for
an equation with quadratic coefficients the new problem generally
is as hard, or harder than the original problem.

    One may also try Laplace's method: write d(D) as the integral
transform of a kernel, then show that solving the original
difference equation is equivalent to solving a differential
equation obeyed by the kernel.  The differential equation
corresponding to \eq{deq} is of a type unknown to me.  It has four
regular singular points. If one attempts to solve the differential
equation with a series solution, the recurrence relation for the
coefficients in the series is as hard to solve as the original
difference equation, \eq{deq}.

    Although analytic solutions are hard to find,
numerical solutions are easy to implement. For low values of the
L's, one may write \eq{recurrD} as a matrix equation,
M$\cdot\vec{c} = \lambda \vec{c}$, and solve for the eigenvalues
of M.  For larger values of the L's, I develop a WKBJ technique in
the next section.  For the rest of this section, I will discuss
exact symmetries of \eq{recurrD}.

    \Eq{recurrD} possesses the following symmetry, which is easy to prove.
Write \eq{recurrD} in a matrix notation, M$\cdot\vec{c} = \lambda
\vec{c}$, and let $\{ c(D;\lambda)\}$ be the components of a
vector $\vec{c}$ which satisfies \eq{recurrD} with eigenvalue
$\lambda$.  Then the vector with components $\{ (-1)^D
c(D;\lambda)\}$ also satisfies the equation, with eigenvalue
$-\lambda$. It follows that the eigenvalues occur in pairs
($\lambda, -\lambda$) , except for possibly the zero eigenvalues.

    This is a good place to mention that the operator needed by
Thiemann \cite{ThiemSN} for his construction of the spin network
formalism is actually the absolute value, $\mid (V_3)^2\mid$ ,
rather than $(V_3)^2$ itself; hence the relevant eigenvalues are
$\mid \lambda_3 \mid$, and the theorem just proven states that
eigenvalues of $\mid (V_3)^2\mid$ are at least doubly degenerate
(except possibly the zero eigenvalues).

    It is also possible to prove that there is at most one zero
eigenvalue, for given values of $L_x,L_y,$ and F, and to construct
the zero eigenfunction  explicitly. When $\lambda = 0$,
\eq{recurrD} collapses from a second order recurrence relation to
a first order relation. (The relation connects every other value
of D, relating c(D+1) to c(D-1); it is first order, with increment
2 rather than 1.) Square $c(D+1)/c(D-1)$ to get rid of the square
roots; one then has a first order equation for the squares of the
c's, with rational coefficients. This is a standard form, with
solution known up to a normalization constant N \cite{MT} . \bea
    c(D) &= & N \sqrt{f(D-1)/f(D)};\nonumber \\
    f(D)&=&
    \left(\frac{L_y-F+D}{2}\right)!\left(\frac{L_y+F-D}{2}\right)!\times\nonumber
    \\
    & &
    \left(\frac{L_x-F-D}{2}\right)!\left(\frac{L_x+F+D}{2}\right)!,\label{0eig}
\eea for D = max D, max D - 2, max D - 4, $\ldots$, min D; and
c(D) = 0  otherwise.

    \Eq{0eig} must satisfy the boundary conditions that c(D) vanishes
outside the limits max D $\geq $D $\geq$ min D.  To find the
limits on D, I note that (although the symmetry is U(1)) I am
using a basis of SU(2) spherical harmonics. Therefore D is
constrained by the SU(2) limits $ - L_i \leq m_i \leq +L_i $ .
These limits may be turned into limits on F$\pm$ D by using the
definitions F $= (m_x + m_y)/2$ ,D $= (m_x - m_y)/2$. It is then
straightforward to derive the following limits on D. \be
    max(-L_x -F,-L_y + F) \leq D \leq min(+L_x -F,+L_y + F)
\ee From these limits, the denominator f(D) in \eq{0eig} becomes
infinite for D = max D + 2, max D + 4,$\ldots$ and for  D = min D
- 2, min D - 4,$\ldots$, which enforces the boundary condition.
Note the numerator is finite at those points.

    Since the recurrence relation connects only every other value of
D, one might suppose there is another zero eigenvalue, with c(D)
given by \eq{0eig} for D = max D -1, max D - 3,$\ldots$, min D +1;
and D = 0  otherwise. However, this solution does not obey the
boundary conditions; e.g. c(max D + 1) is non-zero.

    Note also that the boundary condition requires the series D
= max D, max D - 2, $\ldots$ to terminate at min D, rather than
min D + 1. If the series terminates at min D + 1, then from
\eq{0eig} c(min D - 1) will be non-zero, which violates the
boundary condition. It follows  that min D and max D must differ
by an even integer, and the total number of allowed values of D,
max D - min D + 1, must be an odd integer. This result is
consistent with our earlier result that non-zero eigenvalues
always occur in pairs $(\lambda, - \lambda)$. For given values of
$(L_x,L_y,F)$ and therefore given values of (max D, min D), there
will be one zero eigenvalue if the number of allowed values of D
is odd; otherwise there will be no zero eigenvalues.

    The \eq{recurrD} links c's all having
the same value of the parameter F.  Put another way, there is one
set of equations \eq{recurrD} for each value of F.  A symmetry
relates the equations for F to the equations for -F, however, so
that there is no need to solve both sets of equations.  Let
\{c(D;F)\} be a solution to \eq{recurrD} with eigenvalue
$\lambda$; then \{c(-D;F)\} is a solution to the equations
\eq{recurrD} with F$\rightarrow$ -F and eigenvalue $-\lambda$.
(Temporarily I have restored the suppressed F dependence of the
c's, for clarity.) The proof is straightforward, because changing
(D,F) to (-D,-F) in the coefficients of \eq{recurrD} interchanges
the two terms on the left, therefore changes the sign of the left
hand side. (The c's on the left must be relabeled correctly; for
example, c(D+1) becomes c(-D-1), not c(-D+1).)

\section{WKBJ}

    In this section I use results from a previous paper on WKBJ
solutions to recurrence equations \cite{nevwkbj}, and I obtain a
WKBJ solution for the recurrence relation \eq{recurrD}.  Since
derivations were given in the earlier paper, for the most part I
shall avoid derivations and motivate results using physical
arguments. However, the earlier paper applied the theory to the 6J
symbols, where there is no need to quantize an eigenvalue, hence
no need to derive quantization conditions, formulas analogous to
$\int p dx = (n+1/2)h$ in standard quantum mechanics.  Later in
this section I include some detail from the previous paper, enough
to extend the theory slightly and derive a quantization condition.

    It is perhaps not surprising that the recurrence relation
\eq{recurrD} has a solution of WKBJ type,
\begin{equation}
    c(D)= A \exp(iS). \label{amplph}
\end{equation}
A recurrence relation may be turned into a second order difference
equation, since the c's may be replaced by central differences,
\bea
    \delta^1 c & :=& c(D+1) - c(D);\nonumber \\
    \delta^2 c & := & c(D+1) - 2 c(D) + c(D-1); \nonumber \\
    c(D+1) &=& c(D) + \delta^1 c; \nonumber\\
    c(D-1) &=& -c(D) - \delta^1 c + \delta^2 c.
\eea Differences are very close to derivatives, therefore
recurrence relations are very close to differential equations.

    For a WKBJ solution to be possible, the
recurrence relation does not have to contain the small parameter
$\hbar$.  For example, WKBJ has been applied to the classical
equations describing waves moving through an inhomogeneous medium,
\be
    d^2\psi/dx^2 + k(x)^2\psi = 0 .
\ee A necessary condition for validity of WKBJ in this classical
context is small derivatives, $d^n k/dx^n = $order $k L^{-n}$,
where L is a (large) length characterizing the rate of variation
of the dielectric constant. In the present application,
derivatives are replaced by differences; the quantities analogous
to k are the square root functions in the recurrence relation
\eq{recurrD}; and the large parameter(s) L are the quantities $L_x
\pm F, L_y \pm F $. There is no $\hbar$ , but we are definitely in
the limit of large quantum numbers.

    I now state the requirements for a second order recurrence relation
 \be
    g_-(D,L) ~e(D-1) + g_0(D,L) ~e(D) + g_+(D,L) ~e(D+1) = 0
    \label{stndform}\ee
to have  a WKBJ solution \cite{nevwkbj}.  The parameters L must be
large and the coefficients g must satisfy the following
conditions.\bea
    (g_+ - g_-)/g_- &\leq &\mbox{order}~1/L;\label{condn1}\\
    g_0/g_{\pm} &\leq& \mbox{order unity};\label{condn2}\\
    \delta^n g /g &\leq& \mbox{order}~L^{-n}.\label{condn3}\eea

    Consider first \eqs{condn1}{condn2}.  If these conditions are
not satisfied initially, often a change of independent variable
will lead to g's which satisfy these conditions.    The
coefficients in \eq{recurrD} do not satisfy \eq{condn1}, but I can
remedy this by changing the dependent variable, \be
        c(D) = (i)^D e(D).\label{edef}\ee

This brings \eq{recurrD} to the standard form \eq{stndform}, with
\bea
    g_-(D) &=& \sqrt{(L_x-F-D+1)(L_x+ F +
            D)}\nonumber \\
            &\times&\sqrt{(L_y+F-D+1)(L_y-F+D)};\nonumber \\
    g_+(D) &=& g_-(D+1); \nonumber \\
    g_0(D) &=& -2 \lambda \label{gdef} \eea

    Next consider the third condition, \eq{condn3}.  This
is the analog of the "small derivatives of k" requirement and is
the most important condition.  It implies that the coefficients
behave like polynomials under differencing, rather than like
sinusoids, say.

    To test whether \eq{condn3} is satisfied, it is convenient
to approximate differences by derivatives.  If derivatives are
falling off as 1/L, then so are the differences. \be
    \delta^1 g  \approx (dg/dD) (\Delta D = 1) + order(d^2g/dD^2).
                 \label{diffisderiv} \ee
When a square root in \eq{gdef} is differentiated with respect to
D, in effect the square root is divided by factors of order
(L$\pm$F) $\pm$ D. Therefore the difference is  down by a factor
of order 1/(L$\pm$F) (not order 1/D; D does not have to be large).

    The $g_{\pm}$ will not obey \eq{condn3} (will not behave like
polynomials under differencing) whenever D is near  zeros of the
square roots, where the $g_{\pm}$ are non analytic and differences
$\simeq$ derivatives can be badly behaved.

    Where are these zeros, and what is their physical interpretation?
At \eq{0eig} I derived limits on D using the SU(2) constraints $ -
L_i \leq m_i \leq +L_i$. I recall these limits here: \be
    max(-L_x -F,-L_y + F) \leq D \leq min(+L_x -F,+L_y + F)\label{Dlimits}
\ee From a comparison of \eqs{gdef}{Dlimits} the zeros of the g's
occur at points where D is approaching the limits imposed by
SU(2).  I will refer to the limits on D given in \eq{Dlimits} as
the SU(2) limits on D, and the corresponding zeros of the g's as
SU(2) zeros. I cannot expect a WKBJ solution to work near the
SU(2) limits, and I will have to check that these values of D
occur inside classically forbidden regions, where the solution is
negligible anyway.  I therefore need to locate the turning points,
values of D where the solution shifts from classically allowed
(sinusoidal) to classically forbidden (exponential), in order to
check that the SU(2) limits are outside the turning points.

    To find the turning points, I apply formulas derived in
\cite{nevwkbj}. In classically allowed regions, where the solution
is sinusoidal, the amplitude and phase in \eq{amplph} are given by
\bea
    A^2& =& \mbox{const.}~[(g_+ + g_- - 2\lambda)(-g_+ +3 g_- +2\lambda)]^{-1/2};\label{A}\\
    \delta^1 S &=& \arccos[(g_- -g_+ +2\lambda)/2g_-]\label{cosS}\\
    &=& 2 \arcsin\sqrt{[(g_- +g_+ - 2\lambda)/(4 g_-)]}. \label{sinS}
\eea When using these formulas for initial orientation, it is
permissible to replace $g_+ + g_- = 2 g_-$, because from \eq{gdef}
$g_+ - g_- = \delta^1 g_-$, which is assumed $\ll g_{\pm}$. With
this replacement, the above formulas simplify to \bea
    A^2& \cong& \mbox{const.}~[(g_-^2 - \lambda^2)]^{-1/2};\label{approxA}\\
    \delta^1 S &\cong& \arccos[\lambda/g_-]\label{approxcosS}\\
    &\cong& 2 \arcsin\sqrt{((g_-  - \lambda)/(2 g_-)}. \label{approxsinS}
\eea For the qualitative discussions in this paper I shall use
these approximate formulas. However, for numerical work one should
use the more exact formulas.  Neglecting the difference between
$g_-$ and $g_+$ is equivalent to setting $L(L+1)\cong L^2$, which
introduces significant errors even when the L's are as large as
10.

    In order to visualize the classically allowed region and the SU(2)
zeros, it is helpful to make a rough plot of the function $g_-^2$
versus D.  From \eq{gdef}, $g_-^2$ is a quartic, therefore the
sketch looks like the usual Mexican hat potential. $g_-^2$ has
four zeros at the four SU(2) limits, \eq{Dlimits}. Because of the
min and max in \eq{Dlimits}, the SU(2) allowed region for D is the
segment on the D axis lying between the two zeros closest to the
center of the hat.  Now draw a horizontal line at a height
$\lambda^2$ above the D axis.  From \eq{approxA}, the turning
points are the two values of D (closest to the center of the hat)
where this line cuts the hat. As $\lambda \rightarrow 0$, these
turning points approach the SU(2) limits and the WKBJ
approximation breaks down.

    There is also a problem as the $\lambda^2$ line approaches the
top of the hat.  Recall the physical interpretation of the quantum
number n in the usual quantum mechanical WKBJ formula $\int p dx =
(n+1/2)h$.   n counts the number of half wavelengths which fit
between the two turning points.  In our case, as $\lambda^2$ grows
large, the two turning points coalesce, and n becomes small;
consequently WKBJ will be inaccurate. The "large" quantum number
is not $\lambda^2$, but $\lambda^2$ - order$(L^4)$, since $g_-^2$
is order $(L^4)$ near the top of the hat.  It is perhaps best to
think of the Mexican hat as an upside-down potential.  The "well"
of the potential is at the top of the hat.

    From \eq{approxsinS}, I must take $\lambda \geq 0$, in order for
the sine to have a zero at the same time as the cosine
\eq{approxcosS} becomes $\pm 1$.  I seem to have lost the negative
$\lambda$ eigenvalues. I can recover them if I use $(-i)^D$
instead of $(i)^D$ in \eq{edef}.  This yields \eq{stndform},
except  $g_0 \rightarrow - g_0$. From \eq{edef} this replaces
$\lambda$ by $-\lambda$ everywhere. I have rederived a theorem
from the previous section: the amplitudes for $\lambda$ and
$-\lambda$ differ by  $(-1)^D$.

    I must now derive a solution near turning points, since
the eigenvalues $\lambda$  are quantized by the usual requirement
that the WKBJ solution connects to exponentially decaying
solutions at left and right turning points. Reference
\cite{nevwkbj} derives the necessary connection formulas.  I
change independent variable,\be
    e(D) = Z(D)/\sqrt{g_-}\label{Zdef}\ee

and insert this form into \eqs{stndform}{edef}. I find that Z
obeys \be
    \delta^2 Z + 2(g_- - \lambda)Z /g_- \cong 0.\label{Zeq}
\ee In deriving \eq{Zeq} I have assumed that I am near $g_-^2 -
\lambda^2 = 0$, but far from the SU(2) zeros of $g_-$, so that I
can neglect higher differences of the $\sqrt{g_-}$ in \eq{Zdef}.

    From the Mexican hat plot, there are two turning points
$D_< $ and $D_>$, to the right and left of the center of the hat.
Consider first the smaller turning point $D = D_<$. I expand \bea
    0& \cong & \delta^2 Z + (D-D_<) Z /k_< ;\nonumber \\
    1/k_< & = & [2d(g_-)/dD](D_<)/ \lambda .
    \label{approxZ}
\eea I have assumed that the zero of the second term in \eq{Zeq}
is linear rather than quadratic.  This means  $1/k_<$ cannot
vanish. $1/k_<$ is essentially the slope of the Mexican hat
potential at the smaller turning point, so is indeed non-zero, and
also positive. At the smaller turning point, and in the
classically allowed region, \bea
    z_< &:=& D-D_< \geq 0;\nonumber \\
    1/k_< & \geq & 0.
\eea $k_<$ is order L, assuming that the $\lambda$ and $g_-$ in
the definition of $k_<$, \eq{approxZ}, are coefficients in the
same recurrence relation, therefore are of the same order in L.

     \Eq{approxZ} is a recurrence relation for the Bessel
and Neumann functions.  The Neumann solution can be discarded,
 because it has exponentially diverging behavior in the unphysical
 region, and I get
 \bea
    e(D) &= & \mid c_< \mid J_{-z_< + 2k_<}( 2 k_<)/\sqrt{g_-}
    \nonumber \\
        &\rightarrow & \mid c_< \mid \cos[-2\sqrt{z_< ^3/k_<}/3 +
        \pi/4]\nonumber \\
        & \times& 1/\sqrt{[k_< z_< - (z_< /2)^2]\pi g_-}.\label{smalltp}
 \eea
 The $\sqrt{g_-}$ comes from \eq{Zdef}.  On the
 second line I have used the Debye asymptotic limit, rather than the usual
 Hankel limit, because both the index and the argument of the
 Bessel function are large \cite{MO}.  The form given is valid for $2k_< \gg
z_< > 3 \sqrt[3]{2 k_< }$, z large but not too large. I am free to
choose the overall phase of the wavefunction, and I have chosen
the constant $c_<$
 to be positive.

    Now consider  the larger turning point, $D =
 D_>$, and again expand
 \bea
     0 & \cong & \delta^2 Z + (D-D_>) Z /k_> ;\nonumber \\
    1/k_> & = & [2d(g_-)/dD](D_>)/ \lambda .
    \label{approxZ2}
 \eea
The slope on the right side of the Mexican hat, $1/k_>$, is now
negative, and so  is $z_> = D - D_>$ in the classically allowed
region. Again, the solution is a Bessel function \bea
    e(D) &= & c_> J_{+z_> - 2k_>}( -2 k_>)/\sqrt{g_-}
    \nonumber \\
        &\rightarrow & c_> \cos[-2\sqrt{(-z_>)^3 /(-k_>)}/3 +
        \pi/4]\nonumber \\
        &\times& 1/\sqrt{[k_> z_> - (z_>
        /2)^2]\pi g_-}\label{largetp}
\eea The constant $c_>$ may have either sign, since I have already
used up all the phase arbitrariness in the solution when I chose
$c_<$ to be positive.

    \eqs{smalltp}{largetp} must be matched up with the WKBJ
solution \eqs{A}{sinS}.   The difference equation \eq{sinS} has
the following exact solution for S. \bea
    S(D) &=& S(D_<)+ \sum_{x=D_<}^D \delta^1 S(x)\nonumber \\
        &=& S(D_<) + \sum 2 \arcsin[\sqrt{(g_- - \lambda)/2 g_-}]
                        \label{exactS}\\
    &\cong& S(D_<)+ \sum \sqrt{[2(g_- - \lambda)/ g_-]} \nonumber\\
    &\cong& S(D_<)+ \sum_{z=0}^z \sqrt{z_</k_<}.\label{Sapprox}
\eea The third line is $\arcsin x \cong x$, valid near turning
points.  The last line uses the expansion of \eq{approxZ}.

    I would like to replace the sum in \eq{Sapprox} by an
integral, then integrate to obtain S (equivalently, replace
\[
\delta ^1 S \rightarrow dS/dD = dS/dz, \] then integrate). It is
not immediately clear I may do this, because the square root in
\eq{Sapprox} is not polynomial-like. However, consider the ratio

\be
    r(z) = \Sigma_{m=0}^z \sqrt{m}/(2 \sqrt{z^3}/3),\label{ratio}
\ee

The numerator  is the sum on the last line of \eq{Sapprox}; the
denominator is the approximation to the sum obtained by replacing
the sum by an integral. For $D-D_<$ = z = 3, r = 1.2; for z = 10,
r = 1.07. In words: as I move away from the turning point, the
integral is dominated by regions where the function is
polynomial-like, and the integral becomes a better approximation
to the sum.  Therefore replacing sum by integral in \eq{Sapprox}
is valid near z = $D-D_<$ (but not too near).  I integrate
\eq{Sapprox} to get S and then substitute into \eq{amplph}:

 \bea
    Re A \exp(iS)&=&  A \cos[\int_{D_<}^D \sqrt{z_</k_<}+
    S(D_<)]\nonumber \\
    &=& A \cos[2\sqrt{ z_<^3/ k_<}/3 +
    S(D_<)].\label{WKBJsmall}
 \eea
 Comparing this to
 \eq{smalltp} I get
 \be
    S(D_<) = -\pi/4.
 \ee

    Now consider the WKBJ solution near the larger turning point $D =
 D_>$.
 \bea
    Re A \exp(iS)&=& A \cos[\sum_{D_<}^D(2~\mbox{arcsin})
    -\pi/4]\nonumber
    \\
      &=& A \cos[\sum_{D_<}^{D_>}(2~\mbox{arcsin}) + \sum_{D_>}^D(2~\mbox{arcsin})
      -\pi/4]\nonumber
    \\
      &=& A  \cos[\sum_{D_<}^{D_>}(2~\mbox{arcsin})+\int_0^{z_>}
      \sqrt{(-z_>)/(-k_>)}-
    \pi/4]\nonumber
    \\
    &=&  A  \cos[\sum_{D_<}^{D_>}(2~\mbox{arcsin}) -2\sqrt{(- z_>)^3/(- k_>)}/3
    - \pi/4].\nonumber \\
    & & \label{WKBJlarge}
 \eea
 "arcsin" denotes the arcsin function from \eq{sinS}.  Comparing
 \eqs{WKBJlarge}{largetp}, I find
  \be
    \int_{D_<}^{D_>} 2\arcsin\sqrt{(g_- +g_+ - 2\lambda)/(4 g_-)}dD =
    (n+1/2)\pi. \label{WKBJeigen}
  \ee
 I have used the more accurate expression, \eq{sinS}, for the arcsin.
 The formula contains n$\pi$, rather than 2n$\pi$, because $c_>$
 can have either sign. Except for the sum and the unfamiliar arcsin,
 the connection formula has the
 usual form.  Again,the turning points are the two roots of  $g_-^2 -
 \lambda^2=0$ which are closest to the center of the Mexican hat.
 (For more accuracy, use \eq{A} and find the two roots of $g_- +g_+ - 2\lambda =
 0$.)

        This is a good point to revisit the case
$\lambda^2 \rightarrow 0$ briefly. If one thinks of the Mexican
hat as an upside-down potential, then, because n is large at
$\lambda^2 \rightarrow 0$, one should expect the form \eq{amplph}
to work well, not badly. However,in the WKBJ approach, one always
solves the equations several times: once away from turning points
and once at each turning point. From \eq{approxZ}, 1/k blows up at
a turning point which is also an SU(2) zero, since $g_-$ has a
square root zero there. The problem as $\lambda^2 \rightarrow 0$
is therefore at the turning points.  They need a more careful
discussion which I do not give here.

    If convenient for numerical purposes, I can replace the
 sum by an integral, as I did when computing S near turning
 points.  The arcsin is monotonic and positive, so that (by the
 same arguments as those used to bound series by integrals in the
 Cauchy integral test for convergence) one can bound the sum above and
 below by two integrals with lower limits differing by one unit. These
 bounds should be fairly tight, if the integral extends over the entire
 classically allowed region of order L. (I
 needed to do numerical work at \eq{ratio}, only because
 the integral was confined to a limited region and near turning points, where
 presumably the integral bounds are least restrictive).

    I attempted to obtain an analytic form for $\lambda$, first replacing sum by
 integral in \eq{WKBJeigen}, then carrying our the integral.  When an
 integrand contains an arcsin of complicated
 argument, usually it is best to integrate by parts, which
 replaces the arcsin by a square root (also complicated, but not
 as bad as the arcsin).
 \bea
    \int_{D_<}^{D_>} 2\arcsin\sqrt{(g_- - \lambda)/2 g_-}dD &=&
    -\lambda \int_{D_<}^{D_>}dD~D ~d(g_-)^2/dD\nonumber \\
    &\div&[2 (g_-)^2\sqrt{(g_-)^2 - (\lambda)^2}]
        \label{ellip}
 \eea
(The integrate by parts surface term vanishes.)  \Eq{ellip} is
reducible to elliptic integrals, but I believe this fact is of
little use for extracting an analytic form for $\lambda$.  The
arguments of the elliptic integrals will be functions of the zeros
of $g_-^2 - \lambda^2$, so that the $\lambda$ dependence will be
implicit.

    As a numerical check on calculations of this section, I solved
the exact difference equation on a computer for the case F = 0 and
$L_x = L_y = L = 5.$  I obtained the eigenvalues \be
    \lambda = 26.6,19.6,13.3,7.99,3.59, \label{matrixvalues}
\ee plus five more which were the negatives of these eigenvalues,
plus one zero eigenvalue.  I then used the FindRoot command on
Mathematica to solve the WKBJ quantization condition
\eq{WKBJeigen}.  FindRoot must be given n, plus two initial
guesses as to the value of $\lambda$; FindRoot then iterates,
Newtonian style, until the program converges to a solution.  I
obtained \be
    \lambda = 26.2,19.3,13.1,7.86. \label{WKBJvalues}
\ee
 The four values correspond to n = 0 (largest) through n = 3
(smallest).   I was unable to compute the n = 4 eigenvalue (the
$\lambda^2 \rightarrow 0$ eigenvalue).  The FindRoot program would
not converge. The agreement between \eqs{matrixvalues}{WKBJvalues}
is surprisingly good for this small value of L.\nopagebreak

\end{document}